\documentclass[9pt,twocolumn,twoside]{opticajnl}
\journal{opticajournal}
\setboolean{shortarticle}{false}
\usepackage{comment}
\usepackage{braket}
\usepackage{float}
\usepackage{placeins}
\usepackage{jabbrv}
\usepackage{hyperref}

\title{Spectral Tailoring of Inhomogeneous Optical Response Using Two-Dimensional Coherent Spectroscopy} 

\author[1]{Pradeep Kumar}
\author[1,*]{Rohan Singh}

\affil[1]{Department of Physics, Indian Institute of Science Education and Research Bhopal, Bhopal 462066, India}
\affil[*]{rohan@iiserb.ac.in}

\begin{abstract}
Controlling the coherent optical response of inhomogeneous ensembles is a key challenge in advancing light–matter interaction engineering. We present a comparative study of two spectral tailoring approaches using two-dimensional coherent spectroscopy (2DCS): the prepulse and double-pulse (DP) methods. In the prepulse scheme, a high-intensity pulse induces Rabi oscillations, modulating the 2D spectral amplitude and lineshape when its spectral 
bandwidth matches the ensemble’s full width at half maximum (FWHM). To overcome this limitation, the DP method employs variable inter-pulse delay to generate predetermined periodic spectral modulation without bandwidth constraints. Moreover, tuning the relative phase between DP pulses allows selective switching of frequency components, enabling controlled enhancement or suppression of distinct spectral features.
These observations highlight that, while the prepulse approach is constrained by spectral bandwidth, the DP method provides a more versatile and reliable route to manipulate the coherent optical response of inhomogeneous ensembles.
We are hoping
these findings might stimulate further research in optical switching and coherent storage for quantum memory devices using inhomogeneous ensembles.
\end{abstract}

\setboolean{displaycopyright}{false} % Do not include copyright or licensing information in submission.

\begin{document}

\maketitle

\section{Introduction}

Tailoring the coherent nonlinear optical response has been a pivotal area of research in advancing photonic quantum technologies \cite{Li2003}.
High-intensity excitation pulses and advanced pulse-shaping techniques have been extensively utilized to steer quantum systems toward specific target states and to manipulate their optical responses \cite{Patton2005, Silberberg2009}. 
These approaches have enabled the control of quantum state populations, such as Rabi oscillations in single quantum dots (QDs) \cite{Kamada2001},  manipulation of chemical reaction pathways \cite{Levis2001},  engineering of multiphoton transition probabilities \cite{Agarwal1996}, and  selective excitation of specific resonances in dressed \cite{Bai1985,Wollenhaupt2005,Bracht2023} and few-level quantum systems \cite{Wen2013, Meron2025}.

The study of the nonlinear optical response of quantum dots and other homogeneous systems has remained a central focus due to their promising applications in quantum information processing \cite{Stievater2001,Chen2002}. However, the remarkable collective properties of ensembles such as superfluorescence \cite{Raino2018, Bradac2017,  Biliroglu2022} and entanglement  \cite{ Klimov2015, Zarkeshian2017, Mao2021} also attracted considerable research attention.
Recently, there has been significant growing interest in optically addressing of specific frequencies and states within collectively broad, inhomogeneous ensembles \cite{Suzuki2016, Farina2021,Shinbrough2021,O’Sullivan2022}. 
The ability to modulate the spectral response of these inhomogeneous ensembles offers promising applications, including optical switches \cite{ Popov2005, Fang2014}, frequency converters \cite{Raymer2012, Singh2019}, and quantum memory devices \cite{Lvovsky2009, Bensky2012}.
Furthermore, by tailoring the spectral properties of an inhomogeneous ensemble, quantum emitters or absorbers can be synchronized \cite{Lukin2020}. Such synchronization improves the efficiency of quantum memories, enabling coherent storage and retrieval of quantum information \cite{Rastogi2022}.

The ensemble of these quantum emitters typically exhibits substantial inhomogeneous broadening \cite{Moody2011}, resulting in a distribution of optical transition frequencies.
Despite their intriguing properties, controlling and manipulating the coherent effects of nonlinear optical responses in such ensembles remains challenging 
as these effects are often masked by inhomogeneous broadening \cite{Pelzer2012, Kosarev2020}.
Two-dimensional coherent spectroscopy (2DCS) is a powerful technique to address these challenges associated with inhomogeneity in quantum systems.
This technique has proven highly effective in simultaneously
disentangling homogeneous and inhomogeneous broadening.
While 2DCS has been extensively employed for probing inhomogeneous systems, only a few studies have focused on actively manipulating their coherent optical responses. For instance, earlier work demonstrated coherent control of exciton and biexciton states in quantum dot ensembles using 2DCS combined with a prepulse method \cite{Suzuki2016}. This method enabled spectral tailoring of the optical response using a high intensity of prepulse, although it can also introduce excitation-induced dephasing (EID), leading to rapid signal decay \cite{Wang1993}.
To overcome this limitation, pulse-shaping approaches have been explored to achieve spectral tailoring within the perturbative regime \cite{Meshulach1999,Lim2011,Wen2013, Arkhipov2020,kappe2023,Meron2025}. Among them, the double-pulse method combined with 2DCS has proven promising, enabling selective enhancement of resonances in few-level quantum well systems \cite{Wen2013}.
However, a direct comparison between the prepulse and double-pulse methods both aiming to manipulate the coherent optical response of an inhomogeneously broadened ensemble has not yet been performed.

In this study, we demonstrate spectral tailoring of the coherent nonlinear optical response of an inhomogeneous ensemble
using 2DCS combined with prepulse and double-pulse method. We compare and show how these two distinct approaches differently influence and manipulate the coherent nonlinear optical
response of an inhomogeneous ensemble. Moreover, we address the constraints and advantages of these two different approaches. 
In the prepulse method, varying the prepulse strength drives Rabi oscillations that modulate the inhomogeneous coherent response. However, effective control is achieved only when the prepulse spectral bandwidth is comparable to the fullwidth half maximum (FWHM) of the inhomogeneous distribution, imposing a fundamental constraint on its applicability. In contrast, the DP scheme enables precise manipulation entirely within the perturbative regime, using low-intensity shaped pulses that suppress possibility of excitation-induced dephasing \cite{Wachter2002,Wen2013}. By tuning the relative delay and phase between the DP pulses, this method enables predetermined periodic modulation of spectral lineshapes
without requiring the excitation bandwidth
to match the inhomogeneous distribution
 and allows selective switching of arbitrary frequencies within the tailored response. 
A comparison of both approaches highlights
that, while the prepulse method is effective under specific spectral conditions, the DP method offers greater flexibility and
broader applicability for tailoring inhomogeneous coherent optical responses.
\section{Pre-pulse method}
Integrating the prepulse method with 2DCS has been employed to investigate the detuning-dependent Rabi oscillations and to induce the many-body interactions in quantum dot ensembles \cite{Martin2018, Suzuki2018}. In this approach, an initial prepulse with varying intensity interacts with the sample, from the perturbative to the non-perturbative regime, followed by subsequent 2DCS excitation pulses within the perturbative limit. By tuning the prepulse intensity, this technique serves as a simple and effective tool for controlling and manipulating the coherent optical response of the ensemble.
\begin{figure}[ht]
\centering
{\includegraphics[width=\linewidth]{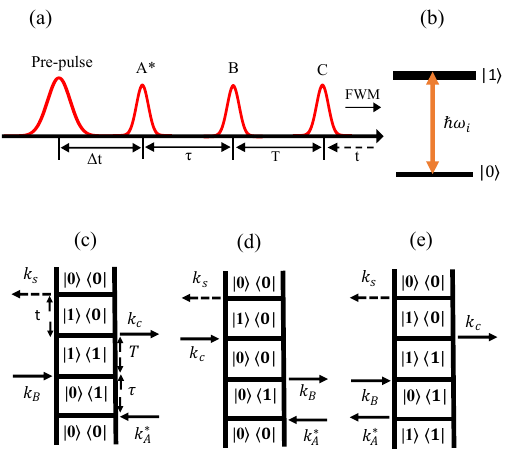}}
\caption{ (a) Optical prepulse excitation sequence is shown here for 2DCS simulation. Delays among the prepule, A*, B, C and the signal are demonstrated as $\Delta$t, $\tau$, $T$, and $t$, respectively.
(b) Energy-level scheme for two-level inhomogeneous ensemble is shown with $\ket{0}$ and $\ket{1}$ representing ground and excited states respectively.
(c)-(e) Possible quantum pathways for two-level system with prepulse excitation are shown using double-sided Feynman diagram.} 
\label{fig:Prepulse_feynman_diag}
\end{figure}
The schematic of the 2DCS pulse sequence with the prepulse excitation of the inhomogeneous ensemble is shown in Fig.~\ref{fig:Prepulse_feynman_diag}(a).
In this approach, the prepulse acts as a pump pulse, first interacting with the sample and coherently preparing the population in both the ground and excited states.
After a delay $\Delta{t}$, three additional rephasing pulses interact with the sample within the perturbative regime.
The rephasing pulse sequence consists of pulses A*, B, and C, where the A* pulse, referred to as the conjugate pulse, arrives first, followed by B and C.
 Delay among these pulses and emitted signal are defined as $\tau$, $T$, and $t$ respectively.
 Nonlinear interaction of these sequence of pulses in the non-collinear geometry with incident wavevector direction 
$\mathbf{k}_A$, $\mathbf{k}_B$, and $\mathbf{k}_C$
give rise third order polarisation, transient four wave mixing signal (TFWM) in phase matching direction $\mathbf{k}_S=-\mathbf{k}_A+\mathbf{k}_B+\mathbf{k}_C$.
\begin{figure}[ht]
\centering
{\includegraphics[width=\linewidth]{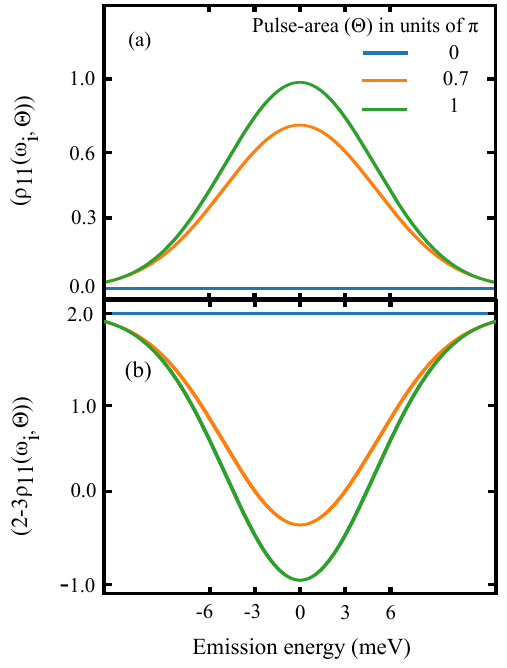}}
\caption{ (a) $\rho_{11}(\omega,\Theta)$ excited population as a function of detuning energy and pulse-area $\Theta$ is demonstrated.
(b) $2-3\rho_{11}(\omega,\Theta)$ variation as a function of detuning energy and pulse-area $\Theta$ is demonstrated.
} 
\label{fig:population}
\end{figure}

We obtain the analytical signal for the prepulse-excitation  by calculating the response function corresponding to the possible double-sided Feynman diagrams shown in Figs. \ref{fig:Prepulse_feynman_diag} (c) - \ref{fig:Prepulse_feynman_diag} (e).
The three quantum pathways illustrated suggest that, for a two-level inhomogeneous ensemble as depicted in Fig. \ref{fig:Prepulse_feynman_diag} (b), the pathway can originate either from the ground state Figs. \ref{fig:Prepulse_feynman_diag} (c) and \ref{fig:Prepulse_feynman_diag}  (d) or from an excited-state population Fig. \ref{fig:Prepulse_feynman_diag} (e).
The analytically calculated transient four-wave mixing (TFWM) signal for a homogeneous two-level system with pre-pulse excitation, incorporating both ground and excited-state population contributions, is given by-
\begin{align}
S_{\text{homo}}^{\text{pre}}(\tau,T,t,w,\Theta) = A 
e^{-i\omega(\tau - t) - (\gamma \tau + \gamma t + \Gamma T)} (2-3 \rho_{11} (\omega,\Theta)), 
\end{align}
where A, represents the signal amplitude, 
$\gamma$ is the dephasing 
rate, and $\Gamma$ denotes the excited-state population decay rate.
The term $\rho_{11}(\omega,\Theta$) corresponds to the excited-state occupation probability and the factor $(2-3\rho_{11}(\omega,\Theta$))
modifies the signal amplitude for a resonance energy $\omega$ influenced by a prepulse with pulse area 
 $\Theta$.
The total TFWM signal for an inhomogeneous two-level system is obtained by incorporating a Gaussian distribution of oscillators, with resonance frequencies $\omega_{i}$ and with $\sigma$ standard deviation for inhomogeneity.
The inhomogeneous TFWM signal is computed by summing the homogeneous TFWM responses for each individual oscillator frequencies, $S_{\text{homo}}^{\text{pre}}(\tau,T,t,w_{i},\Theta)$ and multiplied by their corresponding normalized probability distribution function $g(\omega_{i})$
\begin{align}
S_{\text{inh}}^{\text{pre}}(\tau,T,t,\Theta) = \sum_{i} S_{\text{homo}}^{\text{pre}}(\tau,T,t,w_{i},\Theta) \, g(\omega_{i}).
\end{align}

\begin{figure}[ht]
\centering
{\includegraphics[width=\linewidth]{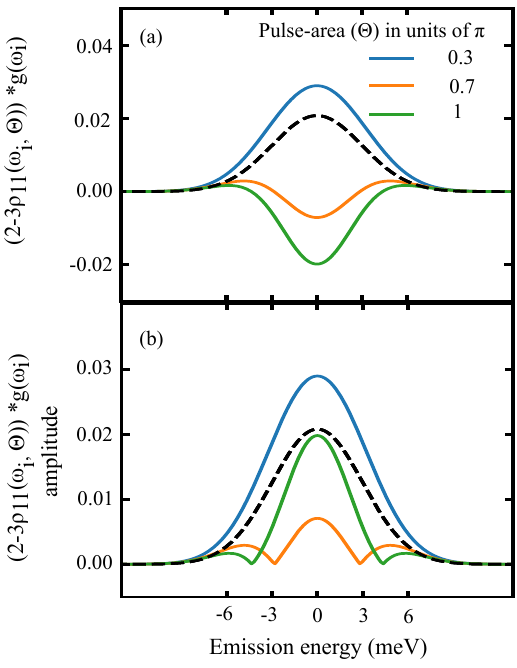}}
\caption{ (a) $2-3\rho_{11}(\omega_{i},\Theta)$ *$g(\omega_{i})$ population weighted by inhomogeneous probability distribution function is demonstrated with changing pulse-area $\Theta$.
(b) Magnitude of modulated population variation as a function of pulse-area $\Theta$ is demonstrated.
} 
\label{fig:modulated_population}
\end{figure}

Equation~(2) for the inhomogeneous TFWM signal comprises a prefactor
$2 - 3\rho_{11}(\omega_i,\Theta)$, which provides direct physical intuition
for the emergence of minima and maxima in frequency-resolved
two-dimensional spectral line shapes. The signal is weighted by this
multiplicative prefactor, which depends explicitly on both the transition
frequency $\omega_i$ and the pulse area $\Theta$. By varying the pulse
area, this prefactor determines whether the contribution from a given
frequency component is enhanced or suppressed, thereby allowing one to
predict the locations of maxima and minima in the frequency-resolved
signal.
For a fixed pulse area, the frequency-dependent response can be analyzed
to identify frequencies $\omega_i$ at which the prefactor
$2 - 3\rho_{11}(\omega_i,\Theta)$ vanishes or attains its extrema.
Specifically,
\begin{equation}
2 - 3\rho_{11}(\omega_i,\Theta) = 0
\quad \Rightarrow \quad
\rho_{11}(\omega_i,\Theta) = \frac{2}{3},
\label{eq:min_condition}
\end{equation}
which corresponds to a minimum in the signal, while
\begin{equation}
2 - 3\rho_{11}(\omega_i,\Theta) = 1
\quad \Rightarrow \quad
\rho_{11}(\omega_i,\Theta) = \frac{1}{3},
\label{eq:max_condition}
\end{equation}
corresponds to a maximum. Using these conditions, one can predict the
appearance of dips or peaks at specific frequencies within the inhomogeneous distribution.

We analyze and demonstrated the response of an inhomogeneously broadened ensemble to excitation by a single optical pulse with changing pulse-strength before the discussion of 2D spectral response.
Figure~\ref{fig:population}(a) shows the frequency resolved excited state occupation population $\rho_{11}(\omega_i,\Theta)$ with changing pulse-area. 
For $\Theta = 0$, the pulse does not induce any population 
transfer and the ensemble remains predominantly in the ground state, resulting 
in $\rho_{11}(\omega_{i},\Theta) \approx 0$ across the entire inhomogeneous 
distribution. As the pulse area increases to 0.7 $\pi$, the pulse excites all the member of ensembles, producing a pronounced peak in $\rho_{11}(\omega_{i},\Theta)$
at  zero detuning. For a $\pi$-pulse, the population inversion is maximized for 
resonant emitters, while emitters far from resonance remain weakly populated due 
to detuning-limited coupling to the prepulse.
Figure~\ref{fig:population}(b) shows the corresponding behavior of the factor 
$2 - 3\rho_{11}(\omega_{i},\Theta)$, which enters multiplicatively in the TFWM signal 
expression. As $\rho_{11}(\omega_{i},\Theta)$ increases with pulse area, this factor 
decreases and eventually becomes negative near resonance for sufficiently strong excitation pulse strength. Physically, this sign change reflects a transition from a 
ground-state-dominated response to an excited-state-dominated response, 
indicating that the intense pulse inverts the population.

Moreover to connect the frequency-resolved population to an experimentally observable 2D spectral lineshapes we observe a single-pulse effect on the ensemble
weighted population while keeping excitation pulse spectrum centered to 
inhomogeneous distribution.
 The multiplication of
the factor \(2 - 3\rho_{11}(\omega_i,\Theta)\) with the probability function of Gaussian inhomogeneous
distribution \(g(\omega_i)\) produces a spectrally modulated population response
across the ensemble, as shown in
Figs.~\ref{fig:modulated_population}(a) and
\ref{fig:modulated_population}(b).
As a result, the weighted response exhibits a reduced amplitude
compared to the earlier discussed unweighted case of Figs. \ref{fig:population}(a) and \ref{fig:population}(b).
Figure \ref{fig:modulated_population}(a) shows how the pulse-induced population modulation is mapped onto the inhomogeneous distribution.
As changing the pulse area contribution of ground state and excited state pathways changes resulting modulated population behavior across the distribution frequencies. 
For weak excitation, at 0.3 $\pi$ pulse area  the response remains positive for all oscillators
and closely
follows the Gaussian spectral envelope, indicating a predominantly ground-state
contribution across the ensemble.
As the pulse area increases,   negative
contributions for near resonance frequencies occur due to prepulse-induced population
inversion, while leaving
off-resonant components largely unaffected at approximately zero value due destructive interference in between the both populations pathways . 
Figure~\ref{fig:modulated_population}(b) shows the corresponding absolute part of the modulated population of Fig.~\ref{fig:modulated_population}(a).
This part reshape the Fig.~\ref{fig:modulated_population}(a) value from excited state dominated negative response to positive response with increasing pulse-area for near resonance frequencies.
Here, we observe a dip around $3~\mathrm{meV}$ at a pulse area of $0.7\pi$, while upon increasing the pulse area to $\pi$, the dip shifts to approximately $4.6~\mathrm{meV}$. This behavior arises from the conditions discussed earlier and is consistent with the results shown in Fig.~\ref{fig:population}(a) for the frequency-resolved population $\rho_{11}(\omega_i,\Theta)$. For these pulse areas, the population reaches $\rho_{11}=2/3$ at these corresponding frequencies, leading to the observed dips.

\begin{figure*}[ht]
\centering
{\includegraphics[width=0.7\textwidth]{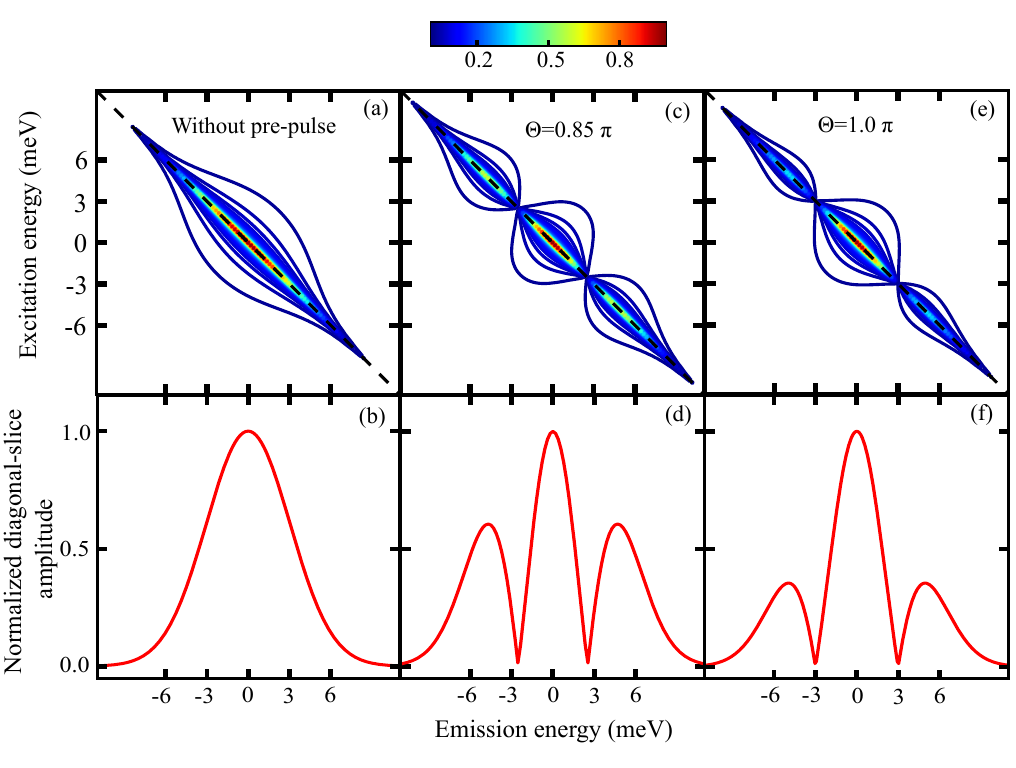}}
    \caption{(a) Absolute part of the 2D spectrum and (b) corresponding diagonal spectral line shape for without pre-pulse, $\Theta = 0 \pi$.
    (c) Modulated 2D spectral amplitude showing Rabi oscillations at a prepulse pulse area of $\Theta = 0.85\pi$, and
    (d) the corresponding modulated diagonal spectral line shape.
    (e) and (f) Rabi-oscillation-induced modulation in the 2D spectral amplitude and the corresponding diagonal spectral line shape observed at a higher prepulse intensity with a pulse area of $\Theta = 1\pi$.}
    \label{fig:prepulse-2Dspectra}
\end{figure*}

We simulated 2D spectra after performing  fourier transform of inhomogeneous TFWM  signal $S_{\text{inh}}^{\text{pre}}(\tau,T,t,\Theta)$ along $\tau$ and t axis while keeping T fixed.
Figure \ref{fig:prepulse-2Dspectra} (a) illustrate the absolute value of the simulated 2D spectra without prepulse .
As anticipated, the resulting spectra resemble the nonlinear interaction of three pulses
with a two-level inhomogeneous ensemble.
The diagonal elongation in the 2D spectra, marked by a dashed line, reflects the finite inhomogeneous broadening due to the Gaussian distribution of oscillator frequencies.
In contrast, the broadening observed along the cross-diagonal direction indicates the homogeneous dephasing, with the dephasing rate set to $\gamma$ =  0.05 meV. 
Normalized  diagonal spectral line shape  extracted from  absloute part of 2D spectra shown in Fig. \ref{fig:prepulse-2Dspectra} (b) has  full-with half-maximum, $\sigma$ = 3 meV. 

Furthermore, to investigate the influence of the prepulse excitation strength, we simulated 2D spectra in the non-perturbative regime.
In the simulations, the prepulse was modeled with a Gaussian temporal envelope having a full width  half maximum
(FWHM) of 150~fs, and a temporal delay of $\Delta t = 3~\mathrm{ps}$ is maintained between the prepulse and the first rephasing pulse.
 Modulation in the 2D spectral amplitude and the corresponding diagonal spectral lineshape are illustrated in Figs. \ref{fig:prepulse-2Dspectra} (c) and \ref{fig:prepulse-2Dspectra} (d) at $\Theta = 0.85\pi$ pulse-area. These modulations in the 2D spectral amplitude arises from coherently induced Rabi oscillations driven by the prepulse.
Consequently
we observe the emergence of three subensemble peaks a central peak at zero detuning energy accompanied by two symmetric side peaks. The ensemble-averaged Rabi splitting energy \cite{Ulhaq2013,Unsleber2015} between the central and side peaks is found to be 4.7 meV for this tailored 2D spectrum and its corresponding diagonal lineshape.
Moreover we find further
increment of the prepulse pulse area up to or beyond $\Theta = \pi$ does not lead to the appearance of additional subensemble peaks or enhanced Rabi oscillations.
While, at this higher prepulse strength, we observe a slight modification in the effective Rabi splitting energy to about 5 meV.
However, due to detuning effects at this higher
prepulse strength,
the amplitude of the subensemble population decreases significantly, as illustrated in Figs. \ref{fig:prepulse-2Dspectra} (e) and in  \ref{fig:prepulse-2Dspectra} (f). 

This behavior arises because emitters that are substantially detuned from the prepulse’s central frequency interact only weakly with the excitation field. As a result, these emitters do not undergo complete population inversion, leading to a suppression of coherent Rabi oscillations across the ensemble \cite{Allen1987}. Consequently, even when the prepulse pulse area is further increased, additional subensemble peaks  or Rabi splitting feautures does not emerge in the 2D spectra.

\begin{figure}[ht]
\centering
{\includegraphics[width=\linewidth]{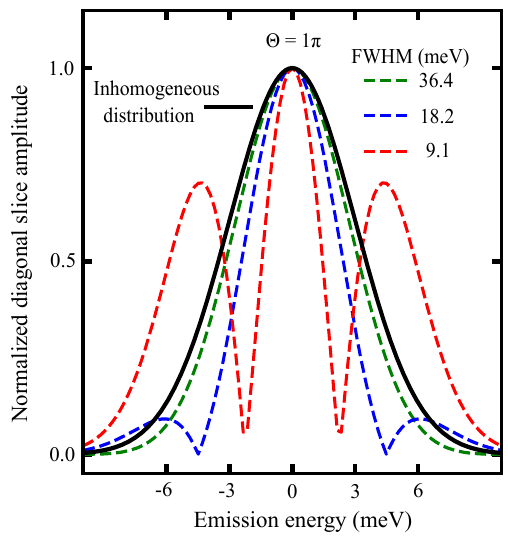}}
\caption{Prepulse spectral-bandwidth-dependent modulation, manifested as Rabi oscillations, is shown in the diagonal lineshape at a fixed prepulse pulse-area of $\Theta = \pi$. Dashed line represents  normalized diagonal spectral lineshapes for varying prepulse spectra and solid line shows the inhomogeneous Gaussian distribution with a full-width half-maximum (FWHM) of 3 meV.}
\label{fig:Diagonal_with_multiple_bw}
\end{figure}

Additionally, we find that manipulation of the coherent optical response using this approach is significantly more effective when the prepulse spectrum is comparable to the spectral bandwidth of the inhomogeneous distribution. 
This relationship plays a key role in tailoring the spectral response of the inhomogeneous ensemble.
We simulated 2D spectra while varying the prepulse spectral bandwidth and, keeping the inhomogeneous distribution full-width  half-maximum (FWHM) constant.
 Diagonal specral lineshapes extracted from the absolute part of the 2D spectral amplitude are shown in Fig. \ref{fig:Diagonal_with_multiple_bw}. Here solid black line represents fixed inhomogeneous distribution, while dashed lines denoted diagonal spectral lineshape amplitude.
These spectral lineshapes highlight the induced Rabi oscillations and corresponding amplitude of the subensemble populations as a function of the prepulse spectral bandwidth.
We observe that increasing the prepulse spectral bandwidth from 9.1 meV to 36.4 meV leads to the disappearance of spectral modulation in the lineshape, accompanied by a decay in the Rabi oscillation amplitude at a fixed prepulse pulse area of $\Theta = \pi$.
This behavior can be explained by considering the definition of the effective Rabi frequency, which depends on both the detuning and the on-resonance Rabi frequency. For a spectrally broader prepulse compared to the inhomogeneous distribution, a larger fraction of emitters is excited off-resonance across the ensemble, leading to a decay in the observed Rabi oscillations due to ensemble-averaged detuning effects \cite{Allen1987,Liu2018}.

%Explanation of this behavior can be understood by effective Rabi frequency definition.The effective Rabi frequency for detuned emitter depend on detuning and on resonance Rabi-frequency. For a spectrally broader prepulse, in comparison to inhomogeneous distribution more emitters are excited off-resonance across the ensemble, explaining the decay of the Rabi oscillations observed in Fig. \ref{fig:Diagonal_with_multiple_bw} due to ensemble average detuning effect.%

\section{Double-pulse method}

In order to circumvent the spectral tailoring constraints associated with the prepulse approach, we implement a double-pulse (DP) method.
It is a pulse-shaping approach for manipulating the coherent optical response of a nanomaterial, through the interplay of time-delay  and relative phase between two temporally separated pulses \cite{Wen2013}. 
By tuning the relative phase between these DP pulses, their contributions can interfere constructively or destructively.
This interference governs the resulting optical response, enabling selective excitation, spectral tailoring, and precise manipulation of quantum state populations.

\begin{figure}[ht]
\centering
{\includegraphics[width=\linewidth]{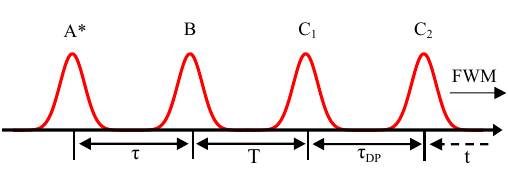}}
\caption{Optical double - pulse excitation sequence is shown here for 2DCS simulation. Final pulse is shaped in to  double pulses $C_{1}$ and  $C_{2}$ . Delays among the pules A*, B,  $C_{1}$,  $C_{2}$ and the signal are demonstrated as $\tau$, $T$, $\tau_{DP}$ and $t$, respectively.}
\label{fig:DP_excitation_pulse_sequence}
\end{figure}

Figure \ref{fig:DP_excitation_pulse_sequence} shows DP method for coherent optical excitation of an inhomogeneous ensemble.
In this approach, we use a rephasing pulse sequence with final excitation pulse split in to double pulses $C_{1}$, and $C_{2}$ with $\tau_{DP}$ delay.
%Delays among these rephasing pulses defined as similar to prepulse approach, with addition to dealy between fianl DP pulses $C_{1}$, $C_{2}$ with $\tau_{DP}$ delay.
Initially three pulses A*, B, $C_{1}$ interact with sample generating a third-order coherence that evolve during  the $ \tau_{DP}$ delay until the arrival of pulse  $C_{2}$.
The signal corresponding to this interaction is emitted in the phase-matching direction $-k_{A}+k_{B}+k_{C_{1}}$, where $k_{A}$, $k_{B}$, $k_{C_{1}}$ represents the wavevector of these incident pulses.
Following the interaction of the final double pulse $C_{2}$ with identical wave vector $k_{C_{2}}$= $k_{C_{1}}$, an additional third-order coherence is generated with delayed by $\tau_{DP}$.
Each pulse in this sequence interacts
with the sample only once. 
Finally total TFWM signal is extracted after nonlinear interactions of these sequence of pulses, with scanning along the detection time delay t.
We consider delta function pulses approximation to calculate the TFWM signal using perturbative approach.
The analytical solution of TFWM signal for DP method is-

\begin{align}
S_{DP}(\tau, T, t) = A1 \left[ e^{i\omega_{01}\tau_{DP} - \gamma\tau_{DP} - \frac{\sigma^2}{2}\left(\tau_{DP}^2 - 2(\tau - t)\tau_{DP}\right)} + e^{i\phi_{DP}} \right], \notag \\
A1 = e^{-i\Delta_{01}(\tau - t) - (\gamma \tau + \gamma t + \Gamma T) - \frac{\sigma^2}{2} (\tau - t)^2}.
\end{align}

\begin{figure}[ht]
\centering
{\includegraphics[width=\linewidth]{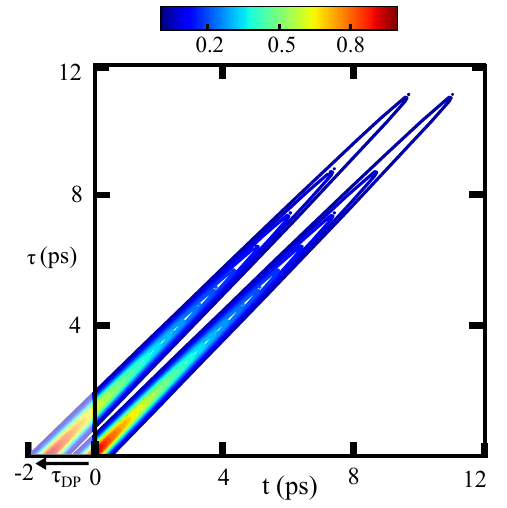}}
\caption{Absolute part of TFWM signal is shown here as two photon-echo seprated by delay $\tau_{DP}$, where $\tau_{DP} = 1.35$ ps.}
\label{fig:TFWM}
\end{figure}

Here, $\Delta_{01} = \omega_L - \omega_{01}$ represents the detuning between the central resonance energy of the excitation pulse ($\omega_L$) and the central resonance energy ($\omega_{01}$) of the two-level inhomogeneous ensemble. The parameter $\phi_{\mathrm{DP}}$ denotes the relative phase between the double pulses. The quantities $\gamma$, $\Gamma$, and $\sigma$ correspond to the dephasing rate, population decay rate, and inhomogeneous linewidth of the ensemble, respectively. The analytically derived final TFWM signal $S_{\mathrm{DP}}(\tau, T, t)$ under DP excitation can be outlined as a combination of two interfering nonlinear TFWM signals, as expressed in equation~(3). The TFWM signal was simulated by scanning along the two time axes $\tau$ and $t$. The relative delay between the double pulses was set to $\tau_{\mathrm{DP}} = 1.35~\mathrm{ps}$, while the relative phase $\phi_{\mathrm{DP}}$ between the pulses was kept at zero. The absolute part of the resulting TFWM signal is shown in Fig. 
\ref{fig:TFWM}, where two photon echoes are observed, separated by 
$\tau_{DP}$
. These photon echoes encapsulate information of both of
the TFWM signal. In Fig. 
\ref{fig:TFWM}, one photon echo appears along the $\tau = t$ line, while the other is shifted along the negative $t$-axis, following the relation $\tau = t + \tau_{\mathrm{DP}}$, and exhibits a reduced signal strength due to dephasing during the delay $\tau_{\mathrm{DP}}$.

Furthermore, we derived the conditions for spectral tailoring by analyzing the  nonlinear optical signal phase behavior  under double-pulse excitation.
Complex signal field of TFWM signal contains phase information of coherent nonlinear optical response.
TFWM
signal field contains constant phase for zero relative delay between double pulses. Consequently Fourier transform of this TFWM signal
field producces constant spectral phase.
However, for nonzero values of $\tau_{DP}$, the Fourier transform of these two temporally delayed, signals introduces a linear spectral phase of the form $\omega{\tau_{DP}}$.
The global spectral phase of the nonlinear signal can therefore be expressed as - 
\begin{equation}
\phi_S = \phi_{\mathrm{DP}} - \omega \tau_{\mathrm{DP}},
\tag{4}
\end{equation} a combination of one a constant uniform phase $\phi_{DP}$, across all the frequencies $\omega$ of the inhomogeneous ensemble , and second, linear spectral phase $\omega{\tau_{DP}}$. 
These phase contributions enable us to determine the conditions for constructive and destructive interference of the nonlinear signal in the spectral domain, from the inference of this $\phi_{s}$. 
Specifically, the signal reaches a maximum when  $\phi_{S}$ = $2n\pi$ and minimum for $\phi_{S}$ = $(2n+1)\pi$, where n is positive integer starting from 0.

%Simulated photon echo is shown in  Fig (2) (a) without considering final pulse as a double pulse. Photon echo signal is simulated for scanning along two time axis t and tau up to 60ps with discrete steps of 1ps.  Decay of this signal is limited by dephasing rate parameters 0.1 meV which correspond to coherence time of this signal up to 6 picosecond. Photon echo is showing elongation along t=$\tau$ direcation represents finite inhomogeneity in our simulation.%

\begin{figure*}[ht]
\centering
{\includegraphics[width=0.6\textwidth]{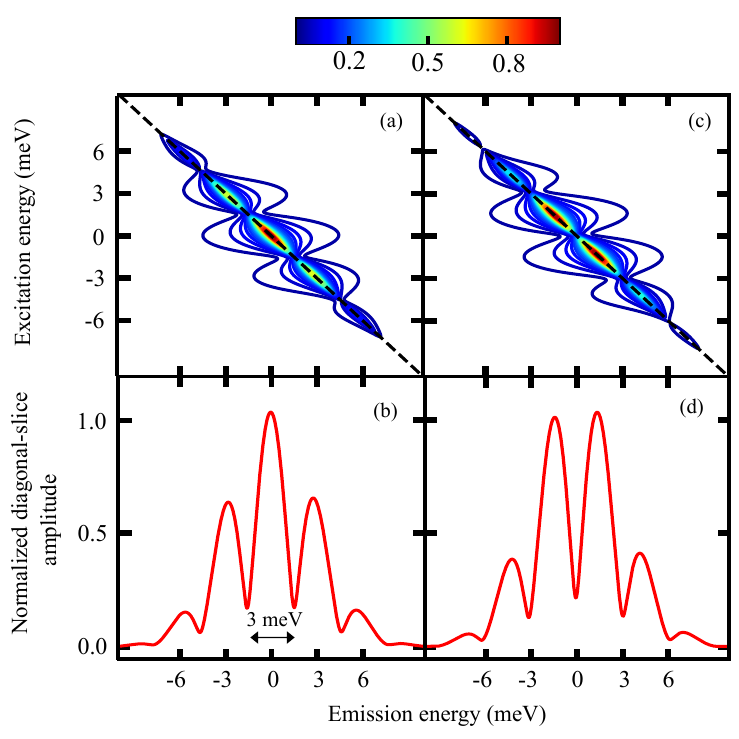}}
\caption{(a) Absolute part of 2D spectra with periodic 2D spectral amplitude of 3 meV and (b) corresponding periodic diagonal spectral line shape at $\tau_{DP}$ = 1.35 ps and $\phi_{DP}$= 0 is shown. The switching behavior at $\omega$ = 1.5 meV energy from destructive minima to constructive maxima is shown through (c) 2D spectral amplitude and (d) corresponding diagonal spectral lineshape.}
\label{fig:all_2D_spectral_lineshapes}
\end{figure*}

Performing Fourier transform of $S_{DP}(\tau, T, t)$ along $\tau$  and \t axes provides the corresponding 2D spectra. Initially, we simulated 2D spectra for $\tau_{DP}$ = 0 and $\phi_{DP}$ = 0 and obtained identical result as illustrated in case of without prepulse in Fig. \ref{fig:prepulse-2Dspectra}  (a) and are therefore not shown.
Moreover 
by controlling the delay between the double pulses, $\tau_{DP}$, we demonstrate that the nonlinear optical response of ensemble spanning a broad frequency range can be precisely tailored. %which was constrained in the prepulse approach.
We produced a predetermined periodic spectral tailoring of subensembles
. 
The enabled spectral modulation of absolute part of 2D spectra and corresponding diagonal spectral lineshape is illustrated in Figs. \ref{fig:all_2D_spectral_lineshapes} (a) and  \ref{fig:all_2D_spectral_lineshapes} (b). 
%We kept relative phase $\phi_{DP}$ = 0 , while vary the $\tau_{DP}$, to produce these spectral lineshapes%.
This observed periodic spectral modulation is governed by the condition $\Delta \omega \tau_{DP}$ = 1, where $\Delta\omega$ represents the periodicity of the spectral amplitude.
%Maxima and minima in these lineshapes are observed at those frequencies where $\omega$$\tau_{DP}$ corresponds to multiples of  $2n\pi$ and $(2n-1)\pi$, respectively.% 
We find, precise tuning of desired spectral amplitude period by chosing appropriate $\tau_{DP}$.
We choose $\tau_{DP}$= 1.35 picosecond, to produce  $\Delta\omega$ = 3 meV periodic spectral amplitude of diagonal lineshape as shown in Fig. \ref{fig:all_2D_spectral_lineshapes} (b), while keeping $\phi_{DP}$=0.
%Althogh a perfect periodic pattern is emerges for destructive interference due to the zero amplitude at the minima of the inhomogeneous distribution. In contrast, the unequal amplitudes at the maxima result in a slightly imperfect periodic pattern for constructive interference.%
Additionally, we note that for producing a narrower and more precise spectral period $\Delta\omega$ requires a longer $\tau_{DP}$ while a broader spectral period is obtained with a shorter $\tau_{DP}$. 
%These results demonstrate the predetermined modulation
%of spectrally resolved response and  overcome the constrained of spectrally resolved tailoring of limited subensembles with in the inhomogeneous distribution using prepulse.% 

%Furthermore, we demonstrate that selective switching of any arbitrary frequency is also possible with in the periodically modulated spectral response. This switching behavior is characterizedby the enhancement or suppression of distinct spectral peaks,governed by interference conditions derived from the frequency-dependent phases of the shaped optical pulse.%

Furthermore, the selective addressing and switching of the tailored nonlinear optical response at arbitrary frequency is demonstrated in  Figs. \ref{fig:all_2D_spectral_lineshapes}(c) and \ref{fig:all_2D_spectral_lineshapes}(d). This switching behavior manifests as the enhancement or suppression of distinct subensemble spectral peaks, governed by interference conditions determined by the frequency-dependent phases of the shaped optical field. 
By applying the conditions for constructive maxima $\phi_{DP} = 2n\pi+\omega_{}\tau_{DP}$
and destructive minima $\phi_{DP} = (2n+1)\pi+\omega_{}\tau_{DP}$, 
the nonlinear response can be selectively switched on or off at specific frequencies.
%we find the ability to selectively switch on and off the nonlinear response of specific frequencies.
%By applying the constructive interference condition $\phi_{\mathrm{DP}} = 2n\pi + \omega \tau_{\mathrm{DP}}$ and the destructive interference condition $\phi_{\mathrm{DP}} = (2n - 1)\pi + \omega \tau_{\mathrm{DP}}$, we achieve selective control over the nonlinear response of specific frequency components.%
As we already discussed periodic spectral response of  Figs. \ref{fig:all_2D_spectral_lineshapes} (a) and \ref{fig:all_2D_spectral_lineshapes} (b)., where one of the spectral minima appears at an energy of $1.5~\mathrm{meV}$ for $\tau_{\mathrm{DP}} = 1.35~\mathrm{ps}$. 
Here we effectively highlights the selective enhancement at $\omega = 1.5~\mathrm{meV}$  energy by applying constructive maxima condition for the 2D spectral amplitude and corresponding diagonal spectral linehape as shown in Figs. \ref{fig:all_2D_spectral_lineshapes} (c) and \ref{fig:all_2D_spectral_lineshapes} (d).
Conversely,  application of the destructive minima condition enable suppression of any selective frequency component.
However, the extent of the suppressed amplitude of any specific frequency is limited by the dephasing rate
of the inhomogeneous ensemble.
%The extent of suppression, however, is limited by the dephasing rate of the inhomogeneous ensemble.%
Moreover, the selective enhancement or suppression of frequencies is not restricted to the only two extreme cases of maxima or minima. 
By carefully tuning the value of $\phi_{\mathrm{DP}}$, the spectral amplitude can be continuously modulated between these limits, enabling precise and flexible control over the nonlinear spectral response. Although these simulations were performed considering the last excitation pulse as the double pulse, the same approach can equivalently be applied when the first conjugate pulse is treated as a double pulse instead.

These spectral tailoring and switching results were obtained by considering the excitation pulses as Dirac delta pulses with infinite spectral bandwidth, while maintaining a finite full width at half maximum (FWHM) for the inhomogeneous distribution. This demonstrates the applicability of the double-pulse (DP) method without the bandwidth constraints associated with the prepulse approach.
However, a key constraint of the DP method is that the delay $\tau_{DP}$
must remain shorter than the coherence time of the sample; otherwise, the spectral tailoring becomes ineffective due to the decay of the coherent signal.

\section{Conclusion \& Outlook}
In summary, we demonstrated spectral tailoring of the coherent nonlinear optical response of an inhomogeneous ensemble using  2DCS. Two approaches are explored and compared: the prepulse and double-pulse (DP) methods combined with 2DCS. The prepulse scheme, which induces Rabi oscillations through a high-intensity prepulse, enables modulation of the 2D spectral amplitude and lineshape but is constrained by the requirement that the prepulse spectral bandwidth should be comparable to the inhomogeneous distribution FWHM. In contrast, in the DP method, signal analytically derived by considering Dirac delta excitation pulses circumvents this limitation, as its effective excitation bandwidth is infinitely broad relative to the finite inhomogeneous width. Consequently, spectral tailoring can be achieved without matching the excitation bandwidth to the ensemble distribution. By optimizing the double pulse delay $\tau_{DP}$  and relative phase $\phi_{DP}$, we demonstrated the predetermined periodic modulation of multiple subensembles and deterministic enhancement or suppression of arbitrary frequencies through phase-controlled interference.
These results are expected to contribute to the development of broadband optical switching and quantum memory technologies using inhomogeneous ensemble.

\bibliography{DP_bib}

@article{Agarwal1996,
  author    = {Agarwal, G. S. and Harshawardhan, W.},
  journal   = {Phys. Rev. Lett.},
  title     = {Inhibition and Enhancement of Two Photon Absorption},
  year      = {1996},
  issn      = {1079-7114},
  month     = aug,
  number    = {6},
  pages     = {1039--1042},
  volume    = {77},
  doi       = {10.1103/physrevlett.77.1039},
  fjournal  = {Physical Review Letters},
  publisher = {American Physical Society (APS)},
}

@Article{Arkhipov2020,
  author    = {Arkhipov, Rostislav and Pakhomov, Anton and Arkhipov, Mikhail and Demircan, Ayhan and Morgner, Uwe and Rosanov, Nikolay and Babushkin, Ihar},
  journal   = {Opt. Express},
  title     = {Selective ultrafast control of multi-level quantum systems by subcycle and unipolar pulses},
  year      = {2020},
  issn      = {1094-4087},
  month     = may,
  number    = {11},
  pages     = {17020},
  volume    = {28},
  doi       = {10.1364/oe.393142},
  fjournal  = {Optics Express},
  publisher = {Optica Publishing Group},
}

@Article{Bai1985,
  author    = {Bai, Y. S. and Yodh, A. G. and Mossberg, T. W.},
  journal   = {Phys. Rev. Lett.},
  title     = {Selective Excitation of Dressed Atomic States by Use of Phase-Controlled Optical Fields},
  year      = {1985},
  issn      = {0031-9007},
  month     = sep,
  number    = {12},
  pages     = {1277--1280},
  volume    = {55},
  doi       = {10.1103/physrevlett.55.1277},
  fjournal  = {Physical Review Letters},
  publisher = {American Physical Society (APS)},
}

@Article{Bensky2012,
  author    = {Bensky, Guy and Petrosyan, David and Majer, Johannes and Schmiedmayer, Jörg and Kurizki, Gershon},
  journal   = {Phys. Rev. A},
  title     = {Optimizing inhomogeneous spin ensembles for quantum memory},
  year      = {2012},
  issn      = {1094-1622},
  month     = jul,
  number    = {1},
  pages     = {012310},
  volume    = {86},
  doi       = {10.1103/physreva.86.012310},
  fjournal  = {Physical Review A},
  publisher = {American Physical Society (APS)},
}

@Article{Biliroglu2022,
  author    = {Biliroglu, Melike and Findik, Gamze and Mendes, Juliana and Seyitliyev, Dovletgeldi and Lei, Lei and Dong, Qi and Mehta, Yash and Temnov, Vasily V. and So, Franky and Gundogdu, Kenan},
  journal   = {Nat. Photonics},
  title     = {Room-temperature superfluorescence in hybrid perovskites and its origins},
  year      = {2022},
  issn      = {1749-4893},
  month     = mar,
  number    = {4},
  pages     = {324--329},
  volume    = {16},
  doi       = {10.1038/s41566-022-00974-4},
  fjournal  = {Nature Photonics},
  publisher = {Springer Science and Business Media LLC},
}

@Article{Bracht2023,
  author    = {Bracht, Thomas K. and Seidelmann, Tim and Karli, Yusuf and Kappe, Florian and Remesh, Vikas and Weihs, Gregor and Axt, Vollrath Martin and Reiter, Doris E.},
  journal   = {Phys. Rev. B},
  title     = {Dressed-state analysis of two-color excitation schemes},
  year      = {2023},
  issn      = {2469-9969},
  month     = jan,
  number    = {3},
  pages     = {035425},
  volume    = {107},
  doi       = {10.1103/physrevb.107.035425},
  fjournal  = {Physical Review B},
  publisher = {American Physical Society (APS)},
}

@Article{Bradac2017,
  author    = {Bradac, Carlo and Johnsson, Mattias T. and Breugel, Matthew van and Baragiola, Ben Q. and Martin, Rochelle and Juan, Mathieu L. and Brennen, Gavin K. and Volz, Thomas},
  journal   = {Nat. Commun.},
  title     = {Room-temperature spontaneous superradiance from single diamond nanocrystals},
  year      = {2017},
  issn      = {2041-1723},
  month     = oct,
  number    = {1},
  volume    = {8},
  doi       = {10.1038/s41467-017-01397-4},
  fjournal  = {Nature Communications},
  publisher = {Springer Science and Business Media LLC},
}

@Article{Chen2002,
  author    = {Chen, Gang and Stievater, T. H. and Batteh, E. T. and Li, Xiaoqin and Steel, D. G. and Gammon, D. and Katzer, D. S. and Park, D. and Sham, L. J.},
  journal   = {Phys. Rev. Lett.},
  title     = {Biexciton Quantum Coherence in a Single Quantum Dot},
  year      = {2002},
  issn      = {1079-7114},
  month     = mar,
  number    = {11},
  pages     = {117901},
  volume    = {88},
  doi       = {10.1103/physrevlett.88.117901},
  fjournal  = {Physical Review Letters},
  publisher = {American Physical Society (APS)},
}

@Article{Fang2014,
  author    = {Fang, Xu and Lun Tseng, Ming and Ou, Jun-Yu and MacDonald, Kevin F. and Ping Tsai, Din and Zheludev, Nikolay I.},
  journal   = {Applied Physics Letters},
  title     = {Ultrafast all-optical switching via coherent modulation of metamaterial absorption},
  year      = {2014},
  issn      = {1077-3118},
  month     = apr,
  number    = {14},
  volume    = {104},
  doi       = {10.1063/1.4870635},
  publisher = {AIP Publishing},
}

@Article{Farina2021,
  author    = {Fariña, Pablo Cova and Merkel, Benjamin and Valencia, Natalia Herrera and Yu, Penghong and Ulanowski, Alexander and Reiserer, Andreas},
  journal   = {Phys. Rev. Applied},
  title     = {Coherent Control in the Ground and Optically Excited States of an Ensemble of Erbium Dopants},
  year      = {2021},
  issn      = {2331-7019},
  month     = jun,
  number    = {6},
  pages     = {064028},
  volume    = {15},
  doi       = {10.1103/physrevapplied.15.064028},
  fjournal  = {Physical Review Applied},
  publisher = {American Physical Society (APS)},
}

@Article{Kamada2001,
  author    = {Kamada, H. and Gotoh, H. and Temmyo, J. and Takagahara, T. and Ando, H.},
  journal   = {Phys. Rev. Lett.},
  title     = {Exciton Rabi Oscillation in a Single Quantum Dot},
  year      = {2001},
  issn      = {1079-7114},
  month     = nov,
  number    = {24},
  pages     = {246401},
  volume    = {87},
  doi       = {10.1103/physrevlett.87.246401},
  fjournal  = {Physical Review Letters},
  publisher = {American Physical Society (APS)},
}

@Article{Kappe2023,
  author    = {Kappe, Florian and Karli, Yusuf and Bracht, Thomas K and Covre da Silva, Saimon Filipe and Seidelmann, Tim and Axt, Vollrath Martin and Rastelli, Armando and Weihs, Gregor and Reiter, Doris E and Remesh, Vikas},
  journal   = {Mater. Quantum Technol.},
  title     = {Collective excitation of spatio-spectrally distinct quantum dots enabled by chirped pulses},
  year      = {2023},
  issn      = {2633-4356},
  month     = jun,
  number    = {2},
  pages     = {025006},
  volume    = {3},
  doi       = {10.1088/2633-4356/acd7c1},
  fjournal  = {Materials for Quantum Technology},
  publisher = {IOP Publishing},
}

@Article{Klimov2015,
  author    = {Klimov, Paul V. and Falk, Abram L. and Christle, David J. and Dobrovitski, Viatcheslav V. and Awschalom, David D.},
  journal   = {Sci. Adv.},
  title     = {Quantum entanglement at ambient conditions in a macroscopic solid-state spin ensemble},
  year      = {2015},
  issn      = {2375-2548},
  month     = nov,
  number    = {10},
  volume    = {1},
  doi       = {10.1126/sciadv.1501015},
  fjournal  = {Science Advances},
  publisher = {American Association for the Advancement of Science (AAAS)},
}

@Article{Kosarev2020,
  author    = {Kosarev, Alexander N. and Rose, Hendrik and Poltavtsev, Sergey V. and Reichelt, Matthias and Schneider, Christian and Kamp, Martin and Höfling, Sven and Bayer, Manfred and Meier, Torsten and Akimov, Ilya A.},
  journal   = {Communications Physics},
  title     = {Accurate photon echo timing by optical freezing of exciton dephasing and rephasing in quantum dots},
  year      = {2020},
  issn      = {2399-3650},
  month     = dec,
  number    = {1},
  volume    = {3},
  doi       = {10.1038/s42005-020-00491-2},
  publisher = {Springer Science and Business Media LLC},
}

@Article{Levis2001,
  author    = {Levis, Robert J. and Menkir, Getahun M. and Rabitz, Herschel},
  journal   = {Science},
  title     = {Selective Bond Dissociation and Rearrangement with Optimally Tailored, Strong-Field Laser Pulses},
  year      = {2001},
  issn      = {1095-9203},
  month     = apr,
  number    = {5517},
  pages     = {709--713},
  volume    = {292},
  doi       = {10.1126/science.1059133},
  publisher = {American Association for the Advancement of Science (AAAS)},
}

@Article{Li2003,
  author    = {Li, Xiaoqin and Wu, Yanwen and Steel, Duncan and Gammon, D. and Stievater, T. H. and Katzer, D. S. and Park, D. and Piermarocchi, C. and Sham, L. J.},
  journal   = {Science},
  title     = {An All-Optical Quantum Gate in a Semiconductor Quantum Dot},
  year      = {2003},
  issn      = {1095-9203},
  month     = aug,
  number    = {5634},
  pages     = {809--811},
  volume    = {301},
  doi       = {10.1126/science.1083800},
  publisher = {American Association for the Advancement of Science (AAAS)},
}

@Article{Lim2011,
  author    = {Lim, Jongseok and Lee, Han-gyeol and Lee, Sangkyung and Ahn, Jaewook},
  journal   = {Phys. Rev. A},
  title     = {Quantum control in two-dimensional Fourier-transform spectroscopy},
  year      = {2011},
  issn      = {1094-1622},
  month     = jul,
  number    = {1},
  pages     = {013425},
  volume    = {84},
  doi       = {10.1103/physreva.84.013425},
  fjournal  = {Physical Review A},
  publisher = {American Physical Society (APS)},
}

@Article{Lukin2020,
  author    = {Lukin, Daniil M. and White, Alexander D. and Trivedi, Rahul and Guidry, Melissa A. and Morioka, Naoya and Babin, Charles and Soykal, Öney O. and Ul-Hassan, Jawad and Son, Nguyen Tien and Ohshima, Takeshi and Vasireddy, Praful K. and Nasr, Mamdouh H. and Sun, Shuo and MacLean, Jean-Philippe W. and Dory, Constantin and Nanni, Emilio A. and Wrachtrup, Jörg and Kaiser, Florian and Vučković, Jelena},
  journal   = {npj Quantum Inf.},
  title     = {Spectrally reconfigurable quantum emitters enabled by optimized fast modulation},
  year      = {2020},
  issn      = {2056-6387},
  month     = sep,
  number    = {1},
  volume    = {6},
  doi       = {10.1038/s41534-020-00310-0},
  fjournal  = {npj Quantum Information},
  publisher = {Springer Science and Business Media LLC},
}

@Article{Lvovsky2009,
  author    = {Lvovsky, Alexander I. and Sanders, Barry C. and Tittel, Wolfgang},
  journal   = {Nat. Photonics},
  title     = {Optical quantum memory},
  year      = {2009},
  issn      = {1749-4893},
  month     = dec,
  number    = {12},
  pages     = {706--714},
  volume    = {3},
  doi       = {10.1038/nphoton.2009.231},
  fjournal  = {Nature Photonics},
  publisher = {Springer Science and Business Media LLC},
}

@Article{Mao2021,
  author    = {Mao, Yingqiu and Gong, Ming and Nemoto, Kae and Munro, William J. and Majer, Johannes},
  journal   = {Appl. Phys. Lett.},
  title     = {Perspective on witnessing entanglement in hybrid quantum systems},
  year      = {2021},
  issn      = {1077-3118},
  month     = sep,
  number    = {11},
  volume    = {119},
  doi       = {10.1063/5.0062842},
  fjournal  = {Applied Physics Letters},
  publisher = {AIP Publishing},
}

@Article{Martin2018,
  author    = {Martin, Eric W. and Cundiff, Steven T.},
  journal   = {Phys. Rev. B},
  title     = {Inducing coherent quantum dot interactions},
  year      = {2018},
  issn      = {2469-9969},
  month     = feb,
  number    = {8},
  pages     = {081301},
  volume    = {97},
  doi       = {10.1103/physrevb.97.081301},
  fjournal  = {Physical Review B},
  publisher = {American Physical Society (APS)},
}

@Article{Meron2025,
  author    = {Meron, Omri and Arieli, Uri and Bahar, Eyal and Deb, Swarup and Ben Shalom, Moshe and Suchowski, Haim},
  journal   = { Light Sci Appl },
  title     = {Shaping exciton polarization dynamics in 2D semiconductors by tailored ultrafast pulses},
  year      = {2025},
  issn      = {2047-7538},
  month     = feb,
  number    = {1},
  volume    = {14},
  doi       = {10.1038/s41377-025-01748-7},
  publisher = {Springer Science and Business Media LLC},
}

@Article{Moody2011,
  author    = {Moody, G. and Siemens, M. E. and Bristow, A. D. and Dai, X. and Karaiskaj, D. and Bracker, A. S. and Gammon, D. and Cundiff, S. T.},
  journal   = {Phys. Rev. B},
  title     = {Exciton-exciton and exciton-phonon interactions in an interfacial GaAs quantum dot ensemble},
  year      = {2011},
  issn      = {1550-235X},
  month     = mar,
  number    = {11},
  pages     = {115324},
  volume    = {83},
  doi       = {10.1103/physrevb.83.115324},
  fjournal  = {Physical Review B},
  publisher = {American Physical Society (APS)},
}

@Article{O’Sullivan2022,
  author    = {O’Sullivan, James and Kennedy, Oscar W. and Debnath, Kamanasish and Alexander, Joseph and Zollitsch, Christoph W. and Šimėnas, Mantas and Hashim, Akel and Thomas, Christopher N. and Withington, Stafford and Siddiqi, Irfan and Mølmer, Klaus and Morton, John J. L.},
  journal   = {Phys. Rev. X},
  title     = {Random-Access Quantum Memory Using Chirped Pulse Phase Encoding},
  year      = {2022},
  issn      = {2160-3308},
  month     = nov,
  number    = {4},
  pages     = {041014},
  volume    = {12},
  doi       = {10.1103/physrevx.12.041014},
  fjournal  = {Physical Review X},
  publisher = {American Physical Society (APS)},
}

@Article{Patton2005,
  author    = {Patton, B. and Woggon, U. and Langbein, W.},
  journal   = {Phys. Rev. Lett.},
  title     = {Coherent Control and Polarization Readout of Individual Excitonic States},
  year      = {2005},
  issn      = {1079-7114},
  month     = dec,
  number    = {26},
  pages     = {266401},
  volume    = {95},
  doi       = {10.1103/physrevlett.95.266401},
  fjournal  = {Physical Review Letters},
  publisher = {American Physical Society (APS)},
}

@Article{Pelzer2012,
  author    = {Pelzer, Kenley M. and Griffin, Graham B. and Gray, Stephen K. and Engel, Gregory S.},
  journal   = {J. Chem. Phys.},
  title     = {Inhomogeneous dephasing masks coherence lifetimes in ensemble measurements},
  year      = {2012},
  issn      = {1089-7690},
  month     = apr,
  number    = {16},
  volume    = {136},
  doi       = {10.1063/1.4704591},
  fjournal  = {The Journal of Chemical Physics},
  publisher = {AIP Publishing},
}

@Article{Popov2005,
  author    = {Popov, A. K. and Myslivets, S. A. and George, Thomas F.},
  journal   = {Phys. Rev. A},
  title     = {Nonlinear interference effects and all-optical switching in optically dense inhomogeneously broadened media},
  year      = {2005},
  issn      = {1094-1622},
  month     = apr,
  number    = {4},
  pages     = {043811},
  volume    = {71},
  doi       = {10.1103/physreva.71.043811},
  fjournal  = {Physical Review A},
  publisher = {American Physical Society (APS)},
}

@Article{Raino2018,
  author    = {Rainò, Gabriele and Becker, Michael A. and Bodnarchuk, Maryna I. and Mahrt, Rainer F. and Kovalenko, Maksym V. and Stöferle, Thilo},
  journal   = {Nature},
  title     = {Superfluorescence from lead halide perovskite quantum dot superlattices},
  year      = {2018},
  issn      = {1476-4687},
  month     = nov,
  number    = {7733},
  pages     = {671--675},
  volume    = {563},
  doi       = {10.1038/s41586-018-0683-0},
  publisher = {Springer Science and Business Media LLC},
}

@Article{Rastogi2022,
  author    = {Rastogi, Anindya and Saglamyurek, Erhan and Hrushevskyi, Taras and LeBlanc, Lindsay J.},
  journal   = {Physical Review Letters},
  title     = {Superradiance-Mediated Photon Storage for Broadband Quantum Memory},
  year      = {2022},
  issn      = {1079-7114},
  month     = sep,
  number    = {12},
  pages     = {120502},
  volume    = {129},
  doi       = {10.1103/physrevlett.129.120502},
  publisher = {American Physical Society (APS)},
}

@Article{Raymer2012,
  author    = {Raymer, Michael G. and Srinivasan, Kartik},
  journal   = {Phys. Today},
  title     = {Manipulating the color and shape of single photons},
  year      = {2012},
  issn      = {1945-0699},
  month     = oct,
  number    = {11},
  pages     = {32--37},
  volume    = {65},
  doi       = {10.1063/pt.3.1786},
  fjournal  = {Physics Today},
  publisher = {AIP Publishing},
}

@Article{Shinbrough2021,
  author    = {Shinbrough, Kai and Hunt, Benjamin D. and Lorenz, Virginia O.},
  journal   = {Phys. Rev. A},
  title     = {Optimization of broadband Λ -type quantum memory using Gaussian pulses},
  year      = {2021},
  issn      = {2469-9934},
  month     = jun,
  number    = {6},
  pages     = {062418},
  volume    = {103},
  doi       = {10.1103/physreva.103.062418},
  fjournal  = {Physical Review A},
  publisher = {American Physical Society (APS)},
}

@Article{Silberberg2009,
  author    = {Silberberg, Yaron},
  journal   = {Annu. Rev. Phys. Chem.},
  title     = {Quantum Coherent Control for Nonlinear Spectroscopy and Microscopy},
  year      = {2009},
  issn      = {1545-1593},
  month     = may,
  number    = {1},
  pages     = {277--292},
  volume    = {60},
  doi       = {10.1146/annurev.physchem.040808.090427},
  fjournal  = {Annual Review of Physical Chemistry},
  publisher = {Annual Reviews},
}

@Article{Singh2019,
  author    = {Singh, Anshuman and Li, Qing and Liu, Shunfa and Yu, Ying and Lu, Xiyuan and Schneider, Christian and Höfling, Sven and Lawall, John and Verma, Varun and Mirin, Richard and Nam, Sae Woo and Liu, Jin and Srinivasan, Kartik},
  journal   = {Optica},
  title     = {Quantum frequency conversion of a quantum dot single-photon source on a nanophotonic chip},
  year      = {2019},
  issn      = {2334-2536},
  month     = apr,
  number    = {5},
  pages     = {563},
  volume    = {6},
  doi       = {10.1364/optica.6.000563},
  publisher = {Optica Publishing Group},
}

@Article{Stievater2001,
  author    = {Stievater, T. H. and Li, Xiaoqin and Steel, D. G. and Gammon, D. and Katzer, D. S. and Park, D. and Piermarocchi, C. and Sham, L. J.},
  journal   = {Phys. Rev. Lett.},
  title     = {Rabi Oscillations of Excitons in Single Quantum Dots},
  year      = {2001},
  issn      = {1079-7114},
  month     = sep,
  number    = {13},
  pages     = {133603},
  volume    = {87},
  doi       = {10.1103/physrevlett.87.133603},
  fjournal  = {Physical Review Letters},
  publisher = {American Physical Society (APS)},
}

@Article{Suzuki2016,
  author    = {Suzuki, Takeshi and Singh, Rohan and Bayer, Manfred and Ludwig, Arne and Wieck, Andreas D. and Cundiff, Steven T.},
  journal   = {Phys. Rev. Lett.},
  title     = {Coherent Control of the Exciton-Biexciton System in an InAs Self-Assembled Quantum Dot Ensemble},
  year      = {2016},
  issn      = {1079-7114},
  month     = oct,
  number    = {15},
  pages     = {157402},
  volume    = {117},
  doi       = {10.1103/physrevlett.117.157402},
  fjournal  = {Physical Review Letters},
  publisher = {American Physical Society (APS)},
}

@Article{Suzuki2018,
  author    = {Suzuki, Takeshi and Singh, Rohan and Bayer, Manfred and Ludwig, Arne and Wieck, Andreas D. and Cundiff, Steven T.},
  journal   = {Phys. Rev. B},
  title     = {Detuning dependence of Rabi oscillations in an InAs self-assembled quantum dot ensemble},
  year      = {2018},
  issn      = {2469-9969},
  month     = apr,
  number    = {16},
  pages     = {161301},
  volume    = {97},
  doi       = {10.1103/physrevb.97.161301},
  fjournal  = {Physical Review B},
  publisher = {American Physical Society (APS)},
}

@Article{Wollenhaupt2005,
  author    = {Wollenhaupt, M. and Präkelt, A. and Sarpe-Tudoran, C. and Liese, D. and Baumert, T.},
  journal   = {Appl. Phys. B},
  title     = {Quantum control by selective population of dressed states using intense chirped femtosecond laser pulses},
  year      = {2005},
  issn      = {1432-0649},
  month     = dec,
  number    = {2},
  pages     = {183--188},
  volume    = {82},
  doi       = {10.1007/s00340-005-2066-0},
  fjournal  = {Applied Physics B},
  publisher = {Springer Science and Business Media LLC},
}

@Article{Wen2013,
  author    = {Wen, Patrick and Nelson, Keith A.},
  journal   = {J. Phys. Chem. A},
  title     = {Selective Enhancements in 2D Fourier Transform Optical Spectroscopy with Tailored Pulse Shapes},
  year      = {2013},
  issn      = {1520-5215},
  month     = may,
  number    = {29},
  pages     = {6380--6387},
  volume    = {117},
  doi       = {10.1021/jp401150d},
  fjournal  = {The Journal of Physical Chemistry A},
  publisher = {American Chemical Society (ACS)},
}

@Article{Zarkeshian2017,
  author    = {Zarkeshian, P. and Deshmukh, C. and Sinclair, N. and Goyal, S. K. and Aguilar, G. H. and Lefebvre, P. and Puigibert, M. Grimau and Verma, V. B. and Marsili, F. and Shaw, M. D. and Nam, S. W. and Heshami, K. and Oblak, D. and Tittel, W. and Simon, C.},
  journal   = {Nat. Commun.},
  title     = {Entanglement between more than two hundred macroscopic atomic ensembles in a solid},
  year      = {2017},
  issn      = {2041-1723},
  month     = oct,
  number    = {1},
  volume    = {8},
  doi       = {10.1038/s41467-017-00897-7},
  fjournal  = {Nature Communications},
  publisher = {Springer Science and Business Media LLC},
}

@Article{Ulhaq2013,
  author    = {Ulhaq, Ata and Weiler, Stefanie and Roy, Chiranjeeb and Ulrich, Sven Marcus and Jetter, Michael and Hughes, Stephen and Michler, Peter},
  journal   = {Opt. Express},
  title     = {Detuning-dependent Mollow triplet of a coherently-driven single quantum dot},
  year      = {2013},
  issn      = {1094-4087},
  month     = feb,
  number    = {4},
  pages     = {4382},
  volume    = {21},
  doi       = {10.1364/oe.21.004382},
  fjournal  = {Optics Express},
  publisher = {Optica Publishing Group},
}

@Article{Unsleber2015,
  author    = {Unsleber, Sebastian and Maier, Sebastian and McCutcheon, Dara P. S. and He, Yu-Ming and Dambach, Michael and Gschrey, Manuel and Gregersen, Niels and Mørk, Jesper and Reitzenstein, Stephan and Höfling, Sven and Schneider, Christian and Kamp, Martin},
  journal   = {Optica},
  title     = {Observation of resonance fluorescence and the Mollow triplet from a coherently driven site-controlled quantum dot},
  year      = {2015},
  issn      = {2334-2536},
  month     = dec,
  number    = {12},
  pages     = {1072},
  volume    = {2},
  doi       = {10.1364/optica.2.001072},
  publisher = {Optica Publishing Group},
}

@Article{Liu2018,
  author    = {Liu, G. and Be’er, O. and Margalit, Y. and Givon, M. and Groswasser, D. and Japha, Y. and Folman, R.},
  journal   = {Phys. Rev. A},
  title     = {Survival of the fittest in the coherent evolution of quantum ensembles},
  year      = {2018},
  issn      = {2469-9934},
  month     = jul,
  number    = {1},
  pages     = {013856},
  volume    = {98},
  doi       = {10.1103/physreva.98.013856},
  fjournal  = {Physical Review A},
  publisher = {American Physical Society (APS)},
}

\end{document}